\documentclass{PoS}
\usepackage{wrapfig,rotating}
\def\be{\begin{equation}}
\def\ee{\end{equation}}

\title{The Effect of Variable Flavour Number Scheme Variations on PDFs and Cross Sections}

\ShortTitle{GM-VFNS Variations}

\author{\speaker{R.S. Thorne},\\
 Department Of Physics and Astronomy, University College London\\
          Gower Place, London, WC1E 6BT, UK\\ 
        E-mail: \email{thorne@hep.ucl.ac.uk}}

\abstract{ I consider variations in the definition of a General-Mass 
Variable Flavour Number Scheme (GM-VFNS) for heavy flavour structure 
functions, both at next-to-leading order
(NLO) and at next-to-next-to leading order (NNLO). I also define 
a new ``optimal'' 
scheme choice improving the smoothness of the transition from one 
flavour number to the next. At both NLO and NNLO I investigate 
the variation of the structure function for a fixed set of parton 
distribution functions (PDFs) 
and also the change in the distributions when a new MSTW-type 
global fit to data is performed for each GM-VFNS. 
At NLO the parton distributions, and predictions using them at hadron 
colliders, 
can vary by $\sim 2\%$ from the mean value. 
Use of the the Zero-Mass
Variable Flavour Number Scheme, which is simpler but only an approximation, 
leads to results a further couple of percent or more outside 
this range. At NNLO there is far more stability with varying GM-VFNS 
definition. Typical changes in 
PDFs and predictions are less than 1$\%$, with most variation at very small
$x$ values. This demonstrates that mass-scheme
variation is an additional and significant source of uncertainty when 
considering parton distributions, but like other theoretical uncertainties,
it diminishes quickly as higher orders are included.}

\FullConference{XVIII International Workshop on Deep-Inelastic Scattering and Related Subjects, DIS 2010\\
		April 19-23, 2010\\
		Firenze, Italy}

\begin{document}

The treatment of heavy flavours, charm 
and bottom, in structure functions has an important 
impact on the PDFs extracted in fits, due to direct data on 
$F_2^h(x,Q^2)$ and also the contribution to the total structure function 
at small $x$. There are two distinct regimes with different descriptions.  
For $Q^2\sim m_h^2$ massive quarks are
created in the final state, and described using the Fixed Flavour Number 
Scheme (FFNS) (see \cite{Laenen:1992zk} for NLO results), 
$
F(x,Q^2)=C^{FF, n_f}_k(Q^2/m_h^2)\otimes f^{n_f}_k(Q^2),
$
where $n_f$ is the number of light quarks. 
This does not sum $\alpha_S^n \ln^n Q^2/m_h^2$ terms in the 
perturbative expansion which may be important. 
At high scales, $Q^2 \gg m_h^2$, heavy quarks behave like 
massless partons and the logs are summed via evolution equations. 
The distributions for different light quark number are related to each other 
perturbatively
$
f^{n_f+1}_j(Q^2)= A_{jk}(Q^2/m_H^2)\otimes f^{n_f}_k(Q^2),
$
where the matrix elements $A_{jk}(Q^2/m_H^2)$, calculated at 
${\cal O}(\alpha_S^2)$ in \cite{Buza:1996wv}, contain the fixed-order 
$\ln(Q^2/m_h^2)$ contributions.
In the $Q^2/m_h^2 \to \infty$ limit 
the description is the Zero-Mass Variable Flavour Number 
Scheme (ZM-VFNS),
$
F(x,Q^2) = C^{ZMVF,n_f}_j\otimes f^{n_f}_j(Q^2).
$
This is approximate, ignoring all ${\cal O}(m_h^2/Q^2)$ 
corrections. To correct this shortcoming and obtain a correct description 
between the two limits of $Q^2\leq m_H^2$ and 
$Q^2\gg m_H^2$, one can use a  General-Mass Variable Flavour Number Scheme 
(GM-VFNS).   

The GM-VFNS can be defined from equivalence of the 
$n_{f}$ flavour and $n_f+1$ flavour descriptions at all 
orders, resulting in 
$
C^{FF,n_f}_k(Q^2/m_h^2) = 
C^{GMVF,n_f+1}_j(Q^2/m_h^2)\otimes A_{jk}(Q^2/m_h^2),
$
e.g. at ${\cal O}(\alpha_S)$ 
\be
C^{FF,n_f,(1)}_{2,hg}(Q^2/m_h^2) = 
C^{GMVF,n_f+1,(0)}_{2, h\bar h}(Q^2/m_h^2)\otimes P^0_{qg}\ln(Q^2/m_h^2)+
C^{GMVF,n_f+1,(1)}_{2,hg}(Q^2/m_h^2),
\label{GMVFNSdef}
\ee
The VFNS coefficient functions tend to the massless limits
as $Q^2/m_h^2 \to \infty$, but 
$C^{GMVF}_j(Q^2/m_h^2)$ is only uniquely defined in this limit. 
One can swap ${\cal O}(m_h^2/Q^2)$
terms between $C^{GMVF,(0)}_{2, h \bar h}(Q^2/m_h^2)$ 
and $C^{GMVF,(1)}_{2,g}(Q^2/m_h^2)$ in Eq. (\ref{GMVFNSdef}), and 
at higher orders, 
leading to various prescriptions 
\cite{Aivazis:1993pi,Thorne:1997ga,Chuvakin:1999nx,Tung:2001mv,
Thorne:2006qt}.
The TR GM-VFNS \cite{Thorne:1997ga} highlighted the freedom in choice, 
and enforced correct kinematics via a quite complicated definition. 
The (S)ACOT($\chi$) prescription \cite{Tung:2001mv} applied the simple choice 
$
C^{GMVF,(0)}_{2, h \bar h}(Q^2/m_h^2,z)\propto \delta(z-x_{\max})$,
which gives $F^{h,(0)}_2(x,Q^2)\propto e_h^2
(h+\bar h)(x/x_{\max}, Q^2)$, where  $x_{\max}=Q^2/(Q^2+4m_h^2)$,
and imposes the threshold $W^2=Q^2(1-x)/x \ge 4m_h^2$. 
This gives the usual limit 
$
C^{ZMVF,(0)}_{2, h \bar h}(z)= \delta(1-z)
$
for $Q^2/m_h^2 \to \infty$. The  TR' scheme \cite{Thorne:2006qt}
adopted this and extensions to higher orders (though uses a 
different multiplicative factor of $Q^2/(Q^2+4m_h^2)$ \cite{Forte:2010ta}). 
However, ACOT-type schemes have used 
the same order of $\alpha_S$ above and below $Q^2=m_h^2$, despite
the fact that FFNS is LO at ${\cal O}(\alpha_S)$ while ZM-VFNS starts at 
zeroth order. Instead the TR' definition uses, for example, at LO 
the ${\cal O}(\alpha_S)$ FFNS result for $Q^2<m_h^2$, and for
$Q^2>m_h^2$
\be
F_2^h(x,Q^2)=\alpha_S(m_h^2) 
C^{FF,n_f,(1)}_{2, hg}(1)\otimes 
g^{n_f}(m_h^2)
+ C^{GMVF,n_f+1,(0)}_{2, h \bar h}(Q^2/m_h^2)
\otimes (h+\bar h)(Q^2),
\ee
i.e. it freezes the higher order $\alpha_S$ term when going upwards 
through $Q^2=m_h^2$.
This difference in choice can be phenomenologically important.  
As an alternative, but ultimately equivalent formulation of a GM-VFNS, 
BMSN \cite{Buza:1996wv} and FONLL \cite{Forte:2010ta}
define a scheme in general terms as 
\be
F^{\rm GMVF}(x,Q^2) = F_2^{\rm FF}(x,Q^2) -
F_2^{\rm asymp}(x,Q^2)+F_2^{\rm ZMVF}(x,Q^2),
\ee
where the second (subtraction) term is the 
asymptotic version of the first, 
i.e. all terms ${\cal O}(m_h^2/Q^2)$ are omitted.
There are differences in exactly how the second and third terms are defined 
in detail in different schemes. 
In the simplest applications the $\alpha_S$ order of 
$F^{\rm FF}_2(x,Q^2)$ at low $Q^2$ is the same as that of  
$F^{\rm ZMVF}_2(x,Q^2)$ as $Q^2\to \infty$. There is a version of 
FONLL which uses one power higher in the FFNS term, but it leads to 
part of the higher order contribution persisting as $Q^2 \to \infty$.

\begin{wraptable}{r}{0.4\columnwidth}
\vspace{-0.4cm}
\begin{tabular}{|l|l|l|l|l|}
\hline
scheme & a & b & c & d \\
\hline
GM-VFNS1& 0 & -1 & 1 & 0 \\
GM-VFNS2& 0 & -1 & 0.5 & 0 \\
GM-VFNS3& 1 & 0 & 0 & 0 \\
GM-VFNS4& 0 & 0.3 & 1 & 0 \\
GM-VFNS5& 0 & 0 &0 & 0.1 \\
GM-VFNS6& 0 & 0 & 0 & -0.2 \\
optimal& 1 & -2/3 & 1 & 0 \\
\hline
\end{tabular}
\vspace{-0.2cm}
\caption{Parameter values for different extreme GM-VFNS definitions.}
\vspace{-0.4cm}
\label{tab:GMVFNSdef}
\end{wraptable}

Ideally one  
would like any GM-VFNS to reduce to exactly the correct order 
FFNS at low $Q^2$ and exactly the correct order (one power of $\alpha_S$
lower) ZM-VFNS as 
$Q^2 \to \infty$. At present none do, but this can easily be rectified.  
Let us return to the TR' version of the GM-VFNS. The obstacle 
is the presence of the frozen term as 
$Q^2 \to \infty$ (which 
depends on the PDFs only at low scales, so is a small effect at 
large $Q^2$). In fact, this is not strictly necessary and one 
can have instead
\be
(m_h^2/Q^2)^{a} \alpha_S^n(m_h^2)\!\sum 
C_{2,i}^{\rm FF}(m_h^2)\!\otimes \!f_i(m_h^2)  
\qquad {\rm or}\qquad  (m_h^2/Q^2)^{a}\alpha_S^n(Q^2)\!\sum 
C_{2,i}^{\rm FF}(Q^2)\!\otimes\! f_i(Q^2).
\ee
Any $a >0$ provides the correct limit, though strictly from 
factorization one should 
have $(m_H^2/Q^2)$ times $\ln(Q^2/m_H^2)$ terms. 
There is also more freedom. One can modify the heavy quark coefficient 
function as long as the $Q^2/m_h^2 \to \infty$ limit is maintained. 
However, since this appears in convolutions for higher order subtraction 
terms, we do not want a complicated $x$ dependence. A simple choice is 
\be
C^{GMVF,(0)}_{2,h \bar h}(Q^2/m_h^2,z)\to  (1+b(m_h^2/Q^2)^c)\delta(z-x_{\max}),
\ee
where again variation in $c$ really mimics $(m_h^2/Q^2)$ with 
logarithmic corrections.  One can also modify the argument 
of the $\delta$-function, similar to the Intermediate-Mass 
IM scheme \cite{Nadolsky:2009ge}, 
\be
\xi = x/x_{\max} \to x \bigl(1+(x(1+4m_h^2/Q^2))^d 4m_h^2/Q^2\bigr),
\ee
so the kinematic limit stays the same, but if $d>0 \,(<0)$ small $x$ is less 
(more) suppressed. The default $a,b,c,d$ are all zero, but can vary, being 
limited by fit quality or {\it sensible} choices. 

\begin{figure}
\vspace{-0.52cm}
\centerline{\hspace{-1.6cm}\includegraphics[width=0.4\textwidth]{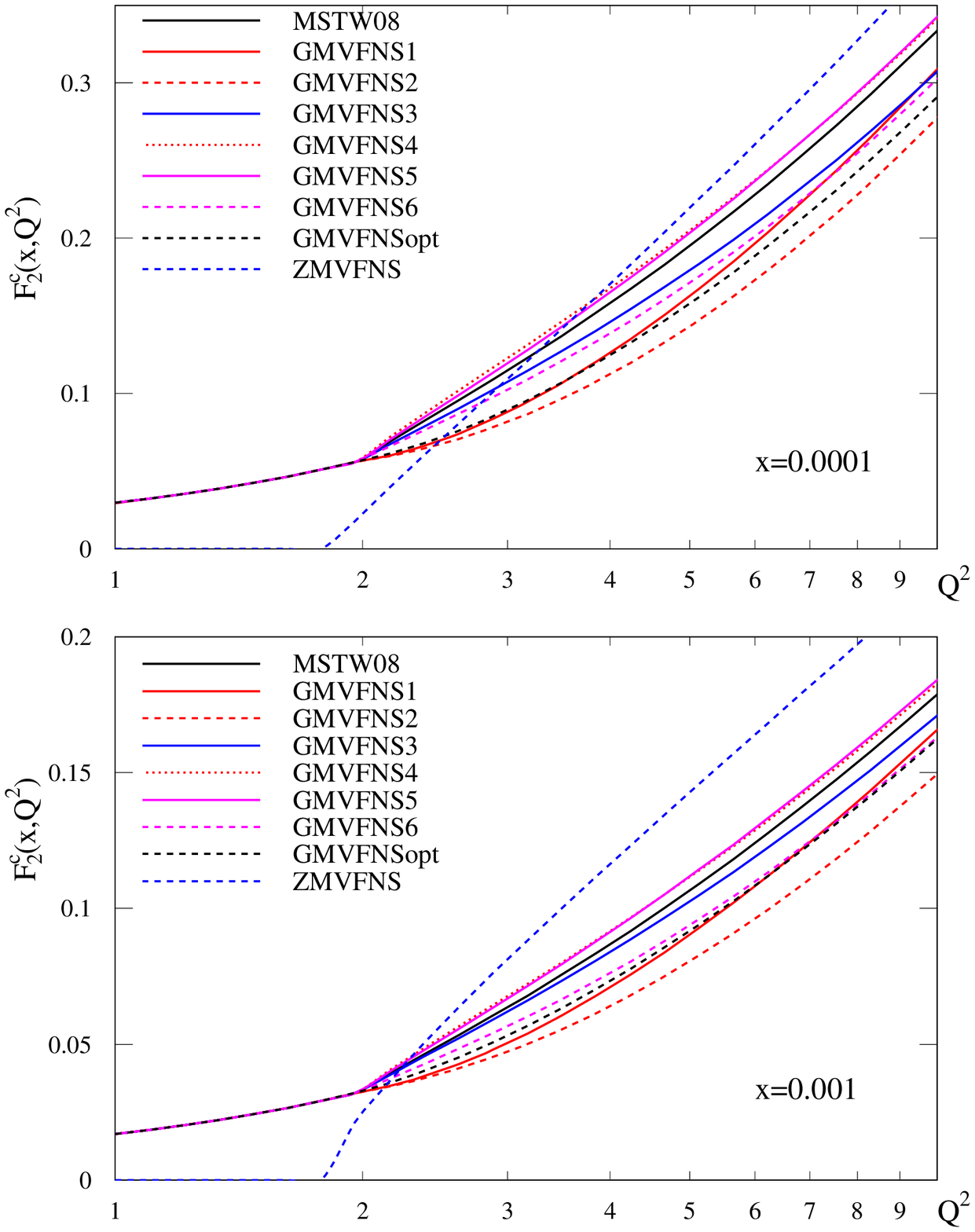}
\includegraphics[width=0.4\textwidth]{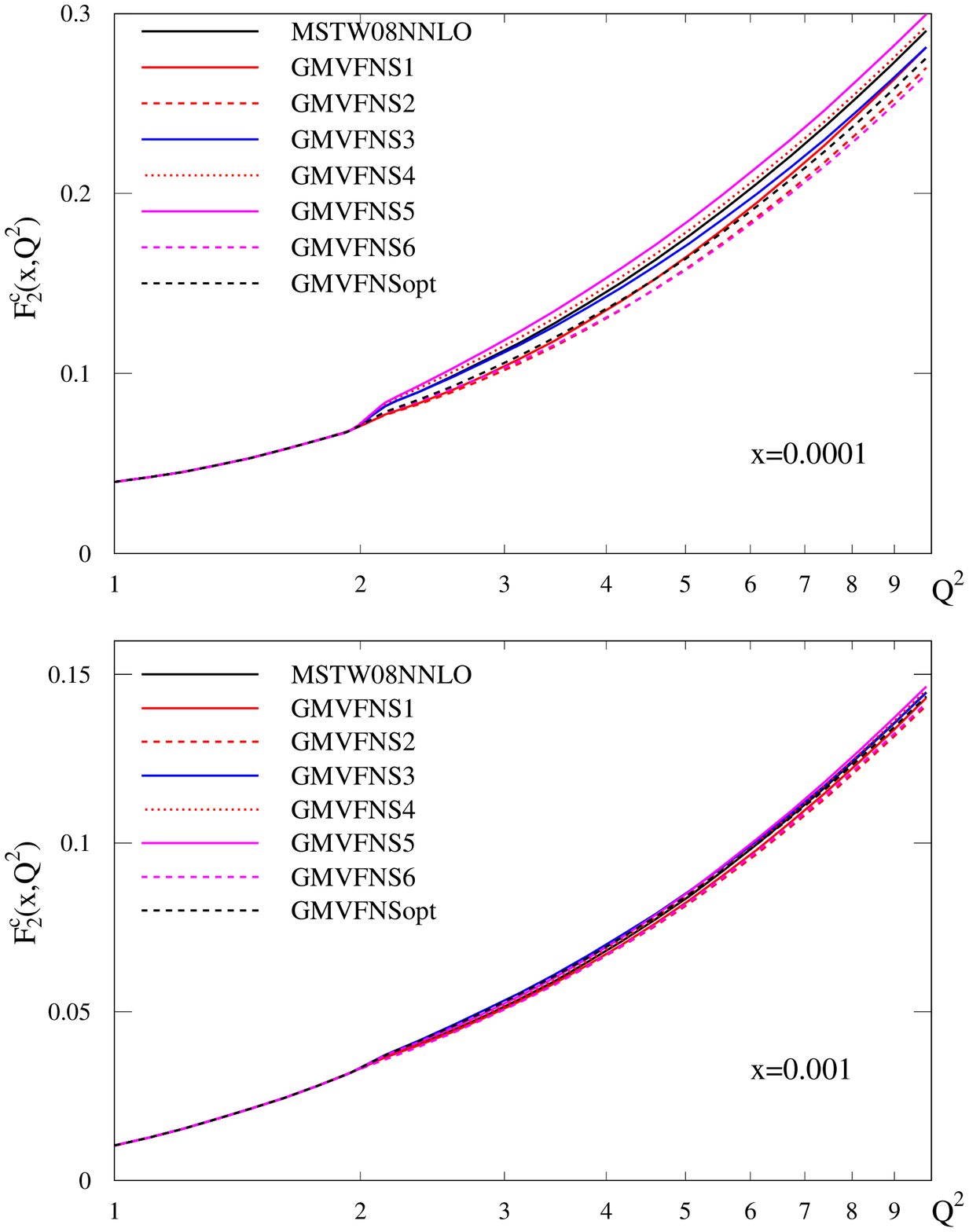}}
\vspace{-1.5cm}
\caption{The variation in $F_2^c(x,Q^2)$ generated from a variety of choices of 
GM-VFNS at NLO (left) and NNLO (right) using the MSTW2008 pdfs in each case.}
\vspace{-0.5cm}
\label{gmvarf2c} 
\end{figure}

\begin{figure}
\vspace{-0.5cm}
\centerline{\hspace{-1.6cm}\includegraphics[width=0.4\textwidth]{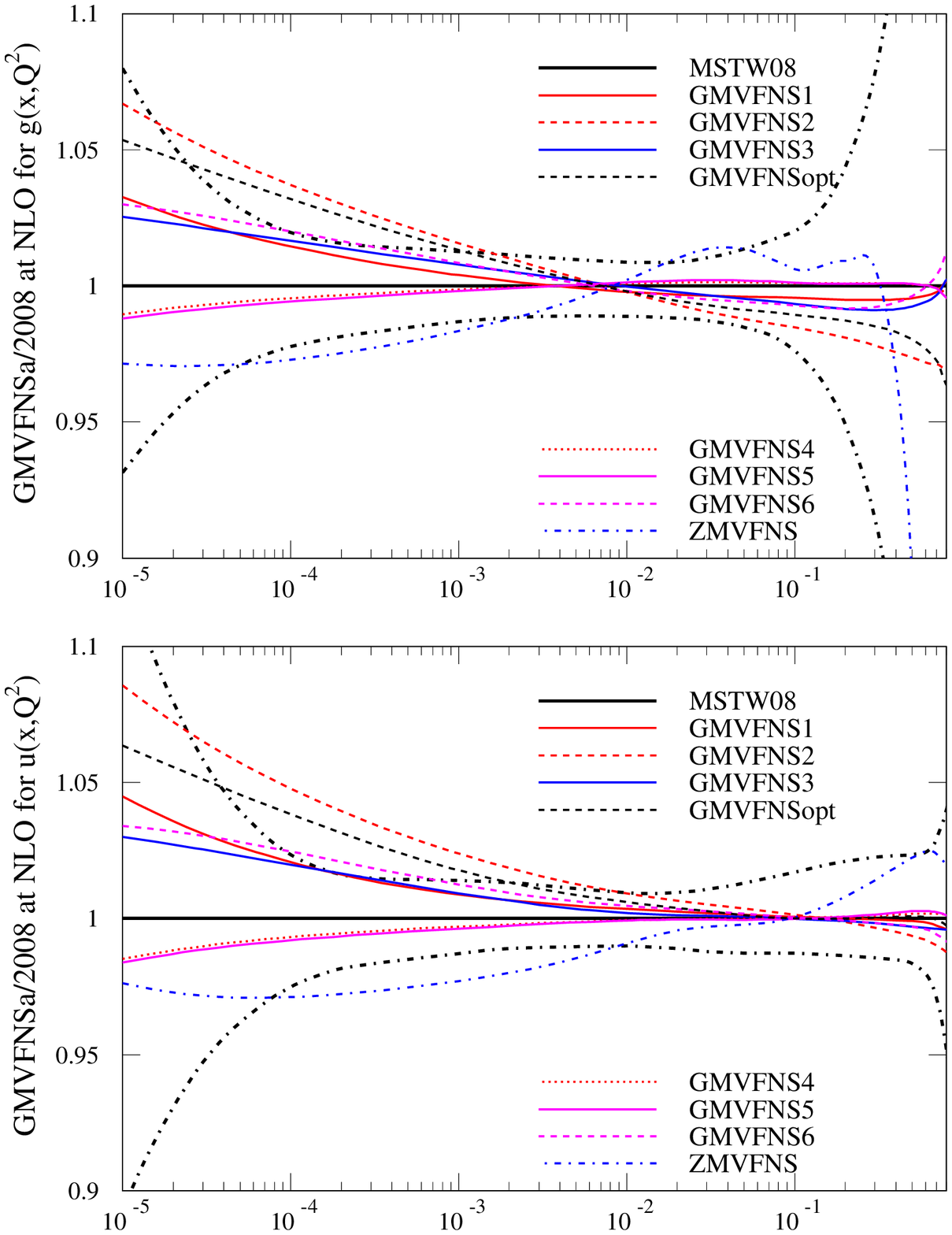}
\includegraphics[width=0.4\textwidth]{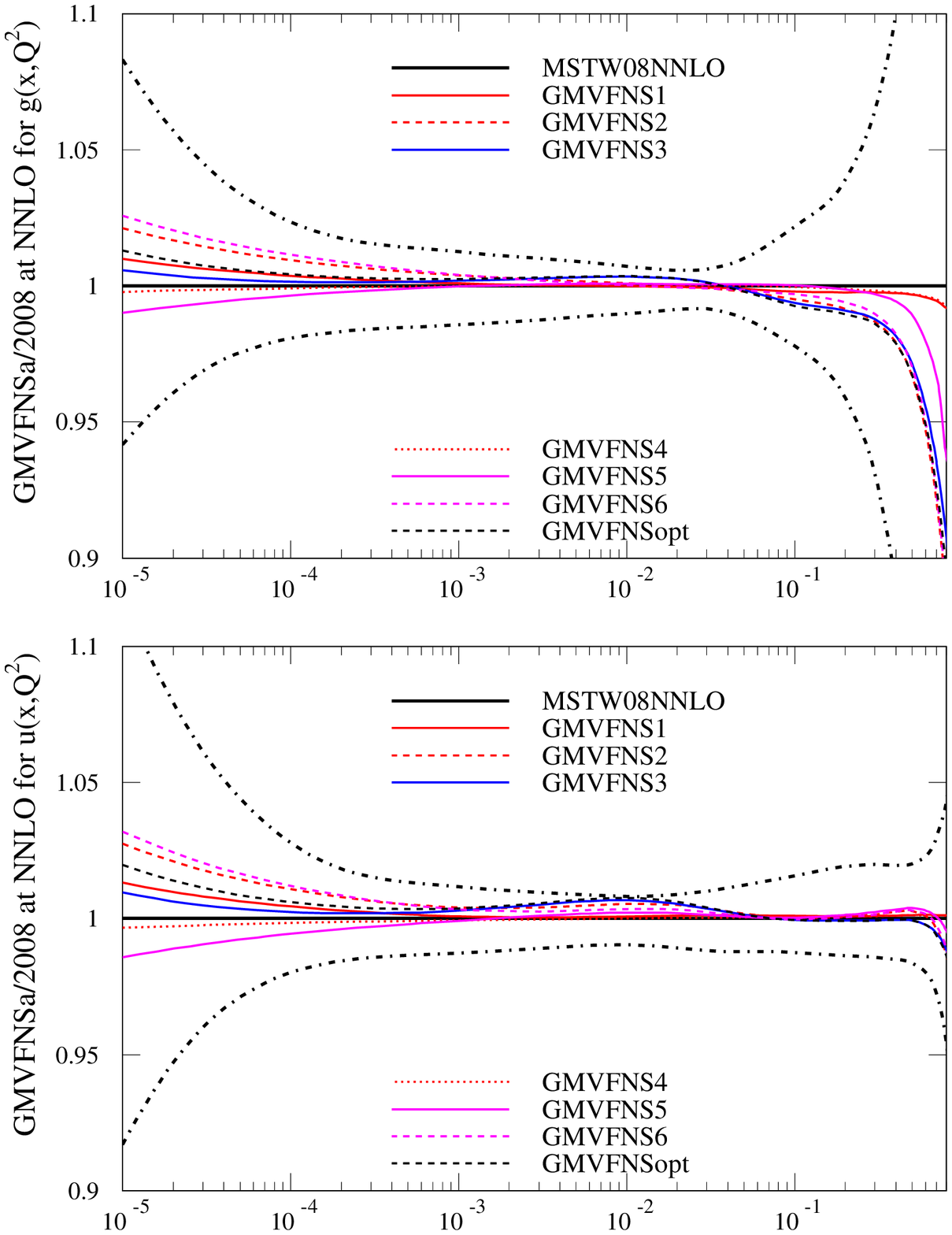}}
\vspace{-1.5cm}
\caption{The variation of PDFs obtained from the best fit from a variety of choices of 
GM-VFNS and the ZM-VFNS at NLO (left) and NNLO (right) as a ratio to the MSTW2008 PDFs.}
\vspace{-0.5cm}
\label{gmvarpart} 
\end{figure}

A variety of different choices defined in Table \ref{tab:GMVFNSdef}
has been tried at NLO and at NNLO along with the ZM-VFNS (at NLO). 
The resulting variations 
in $F_2^c(x,Q^2)$ near the transition point 
due to different choices of GM-VFNS at NLO are shown in the left of 
Fig. \ref{gmvarf2c}. I also define an ``optimal'' scheme which is smooth 
at threshold and reduces to exactly the right limits at high and low $Q^2$. 
There is quite a spread at NLO, though the ZM-VFNS
is far steeper at low $Q^2$ than any GM-VFNS. This spread is  
very much reduced at NNLO, the right of Fig. \ref{gmvarf2c},
with almost zero variation until very small $x$,
showing that NNLO evolution effects are most important in this 
regime.  

\begin{wraptable}{r}{0.51\columnwidth}
\begin{center}
\vspace{-0.8cm}
\hspace{-0.5cm}
\begin{tabular}{|l|ll|ll|}
\hline
PDF set & Tev & & LHC &(${\rm 14~TeV})\!\!\!\!$\\
& $\sigma_Z\,{\rm (nb)}\!\!\!\!\!$  & $\sigma_H$(pb) 
& $\sigma_Z\,{\rm (nb)}\!\!\!\!\!$ & $\sigma_H$(pb) \\
\hline
    $\!\! {\rm MSTW08}$ & 7.207 & 0.7462 & 59.25 & 40.69  \\   
    $\!\! {\rm GMvar1}$ &    $+0.3\%\!\!$ &  $-0.5\%$ &  $+1.1\%\!\!$ & $+0.2\%$   \\    
     $\!\! {\rm GMvar2}$ &  $+0.7\%\!\!$ &  $-1.1\%$ &  $+3.0\%\!\!$ & $+1.5\%$ \\    
     $\!\! {\rm GMvar3}$ &  $+0.1\%\!\!$ &  $-0.3\%$ &  $+1.1\%\!\!$ & $+0.8\%$  \\    
     $\!\! {\rm GMvar4}$ &  $+0.0\%\!\!$ &  $-0.1\%$ & $-0.4\%\!\!$ & $-0.2\%$ \\    
     $\!\! {\rm GMvar5}$ & $-0.1\%\!\!$ &  $-0.1\%$ & $-0.5\%\!\!$ & $-0.3\%$ \\    
     $\!\! {\rm GMvar6}$ &  $+0.3\%\!\!$ &  $-0.4\%$ &  $+1.6\%\!\!$ & $+0.8\%$ \\
     $\!\!{\rm GMvaropt} \!\!\!$& $+0.3\%\!\!$ &  $-1.5\%\!\!$ &  $+2.0\%$ & $+0.4\%$ \\
     $\!\!{\rm Z}$M-V${\rm FNS} \!\!\!$ &  $-0.7\%\!\!$ &  $-1.2\%$ & $-3.0\%\!\!$ & $-3.1\%$ \\
     $\!\!{\rm GMvarcc}$  & $+0.0\%\!\!$ &  $-0.1\%$ &  $+0.0\%\!\!$ & $-0.1\%$ \\    
\hline
    \end{tabular}
\end{center}
\vspace{-0.4cm}
\caption{Predicted cross-sections 
at NLO for 
$Z$ and a 120 GeV Higgs boson at the Tevatron 
and LHC.} 
\vspace{-0.4cm}
\label{cstablenlo}
\end{wraptable}

Global fits are also performed using the same procedure as the MSTW08 fit 
\cite{Martin:2009iq} for all
schemes. At NLO the initial $\chi^2$ for a new scheme can change by $250$, 
but converges to within $20$ of the original. There are improved fits for 
options 1, 3 and 6 and the fit is best for the   
for optimal scheme. The variations in the partons extracted at 
NLO are shown in the left of Fig. \ref{gmvarpart}. The default TR' scheme 
sits near the low end.
Some changes in PDFs exceed the one $\sigma$ {\it uncertainty}.
$\alpha_S(M_Z^2)$ changes by $< 0.0007$ except for the ZM-VFNS 
where it falls by 
0.0015.  The ZM-VFNS PDF is clearly outside than the GM-VFNS band.
For fits at NNLO the initial changes in $\chi^2$ are $< 20$ and
they converge to within $10$ of the original.  
The variations in PDFs extracted at NNLO are shown in the right of
Fig. \ref{gmvarpart}. At worst the changes approach the {\it uncertainty},
but are usually far less. 
Variations in $\alpha_S(M_Z^2)$ are $\sim 0.0003$. 
However, at NNLO the TR' scheme models the ${\cal O}(\alpha_S^3)$ FFNS terms 
at low $Q^2$ 
using leading threshold 
logarithms \cite{Laenen:1998kp} and 
$\ln(1/x)$ terms \cite{Catani:1990eg}.
The latter take the form
$\propto (1-z/x_{\max})^{\tilde a}(\ln(1/z)-{\tilde b})/z,$
where the default is ${\tilde a}=20, {\tilde b}=4$, so ${\tilde a}$ 
and ${\tilde b}$ can be varied. 
Changes in ${\tilde a}$ make little difference. The maximum 
{\it sensible} variation to ${\tilde b}=2$ leads to an effect 
of order the uncertainty at  $x\leq 0.001$.
However, this is largely eliminated if the ${\cal O}(\alpha_S^3)$
contribution dies away, rather than being frozen.    

\begin{wraptable}{r}{0.52\columnwidth}
\begin{center}
\vspace{-0.3cm}
\hspace{-0.6cm}
\begin{tabular}{|l|ll|ll|}
\hline
PDF set & Tev & & LHC &(${\rm 14~TeV})\!\!\!\!$\\
& $\sigma_Z\,{\rm (nb)}\!\!\!\!$  & $\sigma_H$(pb) 
& $\sigma_Z\,{\rm (nb)}\!\!\!\!$ & $\sigma_H$(pb) \\
\hline
     $\!\!{\rm MSTW08}$    & 7.448 & {0.9550}  & 60.93 & 50.51  \\    
     $\!\!{\rm GMvar1}$    & $+0.1\%\!\!$  & $-0.5\%$  & $+0.1\%\!\!$  & $-0.2\%$ \\    
     $\!\!{\rm GMvar2}$    & $+0.3\%\!\!$  & $-0.8\%$  & $+0.5\%\!\!$  & $+0.1\%$\\    
     $\!\!{\rm GMvar3}$    & $+0.4\%\!\!$  & $-0.1\%$  & $+0.5\%\!\!$  & $+0.7\%$ \\    
     $\!\!{\rm GMvar4}$    & $+0.0\%\!\!$  & $-0.2\%$  & $+0.1\%\!\!$  & $-0.1\%$\\    
     $\!\!{\rm GMvar5}$    & $+0.1\%\!\!$  & $-0.3\%$  & $-0.2\%\!\!$ & $-0.2\%$\\    
     $\!\!{\rm GMvar6}$    & $+0.1\%\!\!$  & $-0.9\%$  & $+0.3\%\!\!$  & $-0.2\%$\\
     $\!\!{\rm GMvaropt}$  & $+0.4\%\!\!$  & $-0.2\%$  & $+0.6\%\!\!$  & $+0.8\%$ \\
     $\!\!{\rm GMvarmod \!\!}$  & $-0.2\%$ & $-0.4\%$  & $-1.4\%$ & $-1.0\%$ \\
     ${\!\!\rm GMvarmod' \!\!\!}$ & $+0.0\%$  & $-0.7\%$  & $+0.0\%$  & $+0.1\%$\\
\hline
    \end{tabular}
\end{center}
\vspace{-0.4cm}
\caption{Predicted cross-sections 
at NNLO for 
$Z$ and a 120 GeV Higgs boson at the Tevatron 
and LHC.} 
\vspace{-0.3cm}
\label{cstablennlo}
\end{wraptable}

The predictions for cross-sections are shown 
at NLO in Table. \ref{cstablenlo}.
There is at most a $1.5\%$ variation at the Tevatron.  
There is a $+3\%$ down to $-0.5\%$ variation in $\sigma_Z$ 
at the LHC. The spread in $\sigma_H$ is about halved due to 
the higher average $x$ sampled. The ZM-VFNS is the clear outlier 
in the low direction at the LHC.
GMvarcc denotes variation in the GM-VFNS for  
charged current processes, and clearly the effect is very small indeed.
The predictions at NNLO are seen in Table. \ref{cstablennlo}.  
Other than model dependence -- GMvarmod denotes the variation to 
${\tilde b} =2$ in the ${\cal O}(\alpha_S^3)$ term --
the maximum variations are of order $0.5\%$ at LHC. 
GMvarmod' is when the ${\cal O}(\alpha_S^3)$ terms fall with $Q^2$, and 
also exhibits a very small deviation.

\end{document}